\documentclass[10pt]{IEEEtran}
\usepackage{graphicx}
\usepackage[cmex10]{amsmath}
\usepackage{upgreek}
\usepackage[version=3]{mhchem}
\begin{document}

\title{Superconducting RF Metamaterials Made with Magnetically Active Planar Spirals}


\author{C.~Kurter, A.~P.~Zhuravel, J.~Abrahams, C.~L.~Bennett,
A.~V.~Ustinov, and Steven~M.~Anlage
\thanks{Manuscript received August~3,~2010. This work was supported
by the U.S. Office of Naval Research through Grant No.~N000140811058
and the Center for Nanophysics and Advanced Materials at the
University of Maryland.}
\thanks{C.~Kurter, J.~Abrahams,
C.~L.~Bennett, S.~M.~Anlage are with the Department of Physics,
Center for Nanophysics and Advanced Materials, University of
Maryland, College Park, Maryland 20742-4111 USA (e-mail: ckurter@umd.edu).}
\thanks{A.~P.~Zhuravel is with the B. Verkin Institute for Low Temperature
Physics and Engineering, National Academy of Sciences of Ukraine,
61164 Kharkov, Ukraine.}
\thanks{A.~V.~Ustinov is with the Physikalisches Institut and DFG-Center for
Functional Nanostructures (CFN), Karlsruhe Institute of Technology,
D-76128 Karlsruhe, Germany.}}

\maketitle

\begin{abstract}
Superconducting metamaterials combine the advantages of low-loss,
large inductance (with the addition of kinetic inductance), and
extreme tunability compared to their normal metal counterparts.
Therefore, they allow realization of compact designs operating at
low frequencies. We have recently developed radio frequency (RF)
metamaterials with a high loaded quality factor and an electrical
size as small as $\sim$$\lambda$$/658$, ($\lambda$ is the free space
wavelength) by using Nb thin films. The RF metamaterial is composed
of truly planar spirals patterned with lithographic techniques.
Linear transmission characteristics of these metamaterials show
robust Lorentzian resonant peaks in the sub- 100 MHz frequency range
below the $T_c$ of Nb. Though Nb is a non-magnetic material, the
circulating currents in the spirals generated by RF signals produce
a strong magnetic response, which can be tuned sensitively either by
temperature or magnetic field thanks to the superconducting nature
of the design. We have also observed strong nonlinearity and
meta-stable jumps in the transmission data with increasing RF input
power until the Nb is driven into the normal state. We discuss the
factors modifying the induced magnetic response from single and 1-D
arrays of spirals in the light of numerical simulations.
\end{abstract}


\section{Introduction}
\IEEEPARstart{M}{etamaterials} are engineered structures composed of
 sub-wavelength artificial unit cells that form a
homogeneous effective medium for the electromagnetic waves
propagating through them. This effective response can be described
by a coarse-grained electric permittivity and the magnetic
permeability~\cite{Smith04}. By manipulating these two
electromagnetic parameters, interesting properties unavailable in
nature have been demonstrated, such as negative refractive
index~\cite{Shelby01} and artificial
magnetism~\cite{Wiltshire07,Pendry}.

Most metamaterials are designed to give a magnetic response at
microwave or higher frequency bands~\cite{Shelby01,Chen}. They
usually employ split ring resonators (SRRs), or their derivatives,
and can be generally understood with a lumped-element
inductor-capacitor circuit model~\cite{Shelby01}.

As the wavelength, $\lambda$, of the electromagnetic waves increases
towards the radio frequency (RF) band, miniaturizing the inclusions
forming a metamaterial becomes more
challenging~\cite{WiltshirePSSB07}. The Swiss Roll
metamaterial~\cite{WiltshirePSSB07}, a three dimensional (3-D)
geometry which inspired the SRR design, has achieved a reasonable
magnetic response in the sub-100 MHz band. Since the design is
implemented with normal metals (thick Cu sheets separated by
dielectrics), ohmic losses and bulkiness are the limiting factors.

The most important usage of RF metamaterials reported so far is in
Magnetic Resonance Imaging devices~\cite{Wiltshire01}; however they
can function in a wide variety of applications including
magneto-inductive lenses~\cite{Freire}, RF
antennas~\cite{Ziolkowski}, filters and compact
resonators~\cite{Engheta}.

\section{The Design}

Our Nb RF metamaterials are constructed from micro-structured 2-D
spirals. Earlier experiments~\cite{Baena,Massaoudi} utilized planar
spirals made of thick Cu films ($\sim$35 $\upmu$m thick
in~\cite{Baena} and $\sim$0.25 mm thick in~\cite{Massaoudi}) on
dielectric substrates. Such thick coatings are required to minimize
Ohmic losses, making the design intermediate between 2-D and 3-D.
Further reduction of the resonant frequency of such spirals is not
possible due to the scaling of Ohmic dissipation with spiral length.

Because the electrical size of the elemental spiral is significantly
smaller than the operating wavelength, a quasi-static approach can
be used, and the design can be based on a lumped element analogy as
in SRRs~\cite{Bilotti}. The large number of turns gives a
significant geometrical inductance, $L_g$~\cite{Mohan} and the
spacing between them provides substantial capacitance,
$C$~\cite{Jiang}. Therefore, the geometry can be kept compact enough
to achieve a sub-100 MHz resonance frequency, $f_0 = 1/(2 \pi
\sqrt{(L_g+L_k) C}$ (despite the large $\lambda$). Since the design
is realized with superconducting films, there is an additional term
called kinetic inductance, $L_k$~\cite{Delin} which is closely
linked to the superfluid density.  For example, when the transition
temperature, $T_c$, is approached, the superfluid density goes to
zero, and $L_k$ can make the dominant contribution to the total
inductance, resulting in a significant reduction in $f_{0}$.

\section{Sample Preparation}

The fabrication starts with the deposition of 200 nm Nb thin films
on 350 $\upmu$m thick quartz substrates, 3" in diameter, by the RF
sputtering technique~\cite{Kurter}. The transition temperature of
the film is found to be $T_c \sim 9.25$ K from resistance
measurements taken by a Physical Property Measurement System
(PPMS)~\cite{PPMS}. The main panel of Fig. 1 shows the resistance of
the Nb film for a wide range of temperature from 300 K down to 1.5
K, whereas the top-left inset focuses on the transition region. The
Nb film is patterned into a spiral geometry by photolithography and
reactive ion etching. The etching is performed by a chemically
active mixture of \ce{CF4}/\ce{O2} (10$\%$ \ce{O2}). The wafer is
diced into chips having single spirals and into 1-D arrays composed
of various numbers of spirals. An optical image of one of the single
spirals giving sub-100 MHz resonant features with an outer diameter
$D_o$ of 6 mm and N = 40 turns is shown in the bottom-right inset of
Fig. 1. The width $w$ of the stripes and spacing $s$ between them
are equal and each is 10 $\upmu$m.

\section{Building Blocks of the RF Metamaterial: The Single Spiral}

\subsection{Transmission Data}
The RF transmission measurements are carried out on single elements
of our RF metamaterials in an evacuated cryostat at temperatures
ranging from 300 K down to 4.2 K. The sample is mounted between two
RF magnetic loop probes, 6 mm in diameter, whose axes are parallel
to each other and the Nb spiral, as shown in the inset of Fig. 3.
The spiral resonator is excited by the top probe providing the RF
signal and the transmission is picked up by the bottom probe.  Both
loop probes are connected to an Agilent E5062A RF network analyzer.

\begin{figure}
\centering
\includegraphics[bb=10 174 579 643,width=2.4 in]{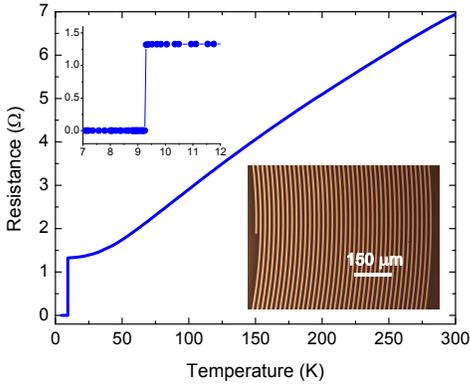}
\caption{(Color online) Main panel: Resistance vs. temperature of
the as-deposited Nb thin film. Top-left inset: Zoom in on the
resistance vs. temperature curve showing the transition in detail.
Bottom-right inset: Optical photograph of a representative Nb spiral
with $D_o$ = 6 mm showing 40 turns in detail.} \label{Fig1}
\end{figure}

The full transmission spectrum on a single spiral with $D_o$ = 6 mm,
N = 40, $w$ = $s$ = 10 $\upmu$m is shown in Fig. 2(a). The data are
taken at $T$ = 4.3 K with an input power of 10 dBm and demonstrate a
fundamental resonance at 74 MHz as well as harmonics located above
100 MHz. The RF current density in the spiral at those resonance
modes are examined with laser scanning microscopy
(LSM)~\cite{Ricci,Zhuravel} in Fig. 2(b)-(f). These images can be
interpreted as showing approximately the local current density
squared flowing in the windings of the spiral (light is large, dark
is small current). The inner and outer borders of the spiral are
marked by dashed circles. In the fundamental mode, an intense
current distribution flows through the middle windings, and gets
weaker near the edges of the spiral (Fig. 2(b)). For the $2^{nd}$
harmonic (Fig. 2(c)), there are two strong current distributions,
presumably of opposite sign, resulting in a partial cancellation of
the fields, and responsible for the smaller transmission amplitude
seen in Fig. 2(a). Higher harmonics (Fig. 2(d)-(f)) show an
increasing number of large-amplitude circles, suggesting that the
spiral acts almost like a distributed circuit resonator carrying
integer multiple half-wavelengths of current in each standing wave
eigenmode.

The transmission data of the same spiral at the fundamental mode for
two different loop-sample distances, $h$, are shown in the main
panel of Fig. 3. Both data sets are taken well below the $T_c$ of
Nb; data with $h$ = 3.25 mm (filled stars) obtained at 4.8 K and
with $h$ = 13.5 mm (empty circles) at 4.4 K. Since Nb is in the
superconducting state, and the driving power is low ($P$ = -10 dBm
for the data shown in Fig. 3), the resonance peaks are independent
of RF input power and are of Lorentzian shape (the solid curves are
fits to the data).

As seen from the figure, larger separation between magnetic loops
leads to sharper transmission peaks, and consequently to a higher
loaded quality factor, $Q_L$, which is calculated as~\cite{Petersan}
$Q_L$ = $f_0$/$\delta f_{(1.5 dB)}$, where $f_0$ is the frequency at
which the peak is located (i.e. resonance frequency) and $\delta
f_{(1.5 dB)}$ is the bandwidth where $|S_{21}|$ is 1.5 dB below its
peak value (see the top-left inset of Fig. 5). The loaded quality
factor of the transmission data taken for $h$ = 3.25 mm is $Q_L
\sim$1502 and for $h$ = 13.5 mm one finds $Q_L \sim$7800. These
quality factor values are up to several orders of magnitude larger
than their normal metal counterparts~\cite{Kurter}. This brings up a
discussion of the effect of coupling between the RF probes on the
transmission $|S_{21}|$ data.

\begin{figure}
\centering
\includegraphics[bb=8 154 596 754, width=2.5 in]{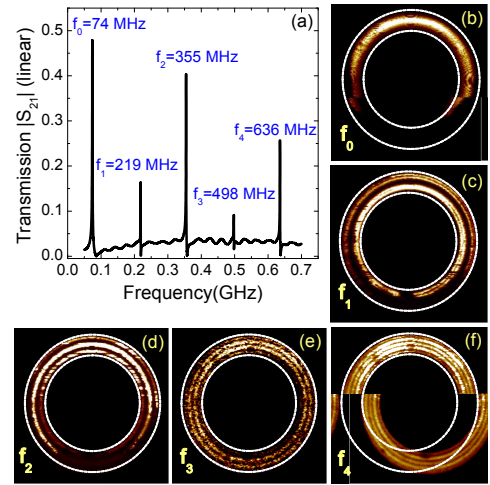}
\caption{(Color online) (a) Measured transmission $|S_{21}|$ vs.
frequency on a single spiral of $D_o$ = 6 mm, $N$ = 40, $w$ = $s$ =
10 $\upmu$m at $T$ = 4.3 K showing the fundamental and higher mode
resonances. (b)-(f) Laser Scanning Microscopy images on the same
spiral showing RF current distributions inside the turns
corresponding to the resonant modes in the transmission data.}
\label{Fig2}
\end{figure}

\subsection{Coupling Between RF magnetic Loops}

The measured $Q_L$ is determined by external and internal loss
factors and can be written as $Q_L$ = $U$$\omega$/$P_t$ where $U$ is
the stored energy in the spiral on resonance, $\omega$ is the
resonant angular frequency and $P_t$ is the total dissipated
power~\cite{Padamsee}. $P_t$ is the sum of power dissipated in the
cavity (Ohmic losses), radiated power, and power leaking out of the
spiral through the coupling loops. Since the measurements are done
well below $T_c$, ohmic and dielectric losses should be the smallest
sources of dissipation. We find that the largest dissipation is due
to the coupling losses from the RF loop probes. To confirm this, we
have simulated the transmission curves numerically on a lossless
single spiral of $D_o$ = 6 mm having the same dimensions as those
used in the experiments shown in Fig. 3, utilizing the High
Frequency Structure Simulator (HFSS)~\cite{HFSS}. As seen in Fig. 4,
$f_0$ shifts to lower frequencies with increasing $h$, which is
consistent with the data in Fig. 3, and the expectation from
electromagnetic perturbation theory for a dominantly magnetic
perturbation~\cite{Ramo}.

The loaded quality factor is observed to increase substantially as
$h$ increases.  This is consistent with an increase in the coupling
quality factor, $Q_{coupling}$ with weaker coupling, and $Q_L^{-1}$
= $Q_0^{-1} + Q_{coupling}^{-1}$. A sigmoidal trend of $Q_L$ is
shown in Fig. 5 along with the polynomial decrease in $f_0$ with
increasing $h$.

\begin{figure}
\centering
\includegraphics[bb=12 152 597 624, width=2.4 in]{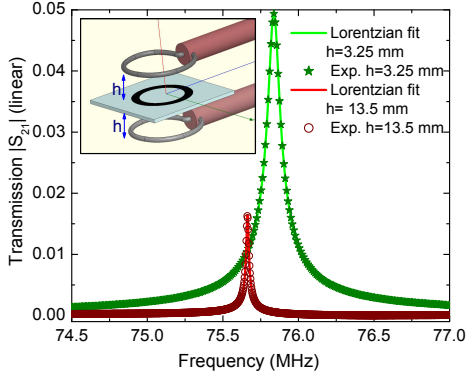}
\caption{(Color online) Main panel: Measured transmission $|S_{21}|$
vs. frequency on a single spiral of $D_o$ = 6 mm for two different
probe distances to the sample, $h$. Filled stars are data obtained
for $h$ = 3.25 mm, and empty circles are those for $h$ = 13.5 mm.
Solid curves are Lorentzian fits, which agree closely with the data,
indicating linear response in both experimental data sets. Inset:
Sketch of the arrangement for the microwave transmission
experiments; a single Nb spiral sandwiched between two magnetic
loops extending from coaxial cables.} \label{Fig3}
\end{figure}

\section{Towards Creation of an Artificial RF Medium: 1-D Arrays of the Spirals}

\subsection{Linear Transmission Data below and above $T_c$}
So far, the discussion has been focused on a single element of the
artificial medium, namely a single spiral showing a sub-100 MHz
resonance. The metamaterial properties should be confirmed by
examining an ordered collection of these elements.

The excitation of the linear array is as follows; the top loop is
aligned with the first spiral of the array whereas the bottom loop
is aligned with the last spiral. The distance between the edges of
two spirals in the array is 1.5 mm, and $h$ is 3.25 mm. The measured
transmission spectrum on a 1-D array of 7 spirals with $D_o$ = 6 mm,
N = 40, $w$ = $s$ = 10 $\upmu$m is shown in the main panel of Fig.
6. The Nb spiral array is excited by an input power of 0 dBm at $T$
= 4.5 K, consequently multiple resonant peaks are observed which
dissapear above the superconducting phase transition at $T_c$. The
inset of Fig. 6 shows the calculated transmission spectrum on the
same spiral array, and gives resonant features at a slightly higher
frequency (HFSS does not include $L_k$ or internal losses in the
calculation).

\begin{figure}
\centering
\includegraphics[bb=12 143 596 622, width=2.4 in]{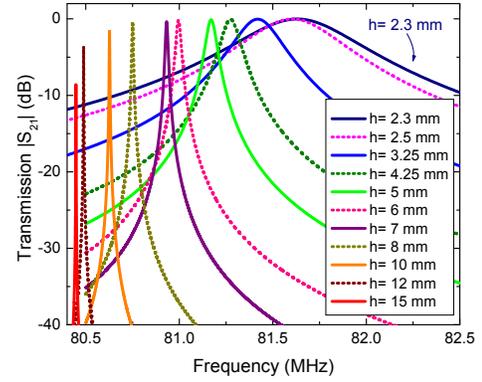}
\caption{(Color online) Calculated transmission $|S_{21}|$ vs.
frequency by HFSS for a set of distances $h$ between a single spiral
of $D_o$ = 6 mm and one of the RF magnetic loops. As the driving
loop is moved away from the spiral, $f_0$ shifts to smaller values
and $Q_L$ becomes larger.} \label{Fig4}
\end{figure}

\begin{figure}
\centering
\includegraphics[bb=12 153 580 621, width=2.4 in]{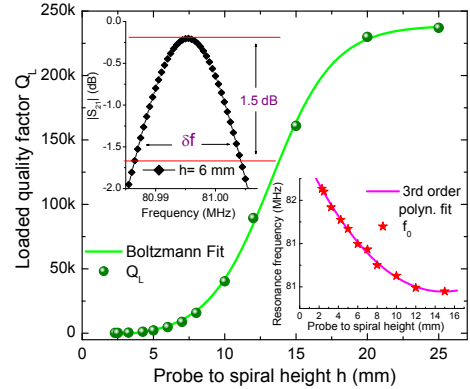}
\caption{(Color online) Main panel: Quality factor vs. driving loop
probe distance to the spiral from HFSS calculations. The data
(filled circles) follow a Boltzmann fit (solid curve).  Top-left
inset: The calculation of $Q_L$ for data at $h$ = 6 mm (filled
diamonds). Bottom-right inset: Resonance frequency decreasing with
increasing probe distance to the spiral (filled stars).}
\label{Fig5}
\end{figure}

\subsection{RF Power Evolution of Transmission Data and Nonlinearity}

In the linear regime, the behavior of transmission data does not
show any significant changes with respect to RF power put into the
spiral cavity. The superconducting phase is strong and RF oscillations
are maintained without significant Ohmic losses (the spiral can be
modeled as an ideal inductor-capacitor (L-C) resonant circuit). As RF power is
ramped up, the magnetic field starts to penetrate the spiral, particularly
at weak spots in the superconducting film~\cite{Ricci}. For
higher RF input power, the
dissipation caused by both vortex dynamics and Joule heating
degrades superconductivity and introduces a resistance $R$ into
the lumped element circuit analogue, and leads to nonlinear resonant
features in the transmission data~\cite{Abdo}.

\begin{figure}
\centering
\includegraphics[bb=12 195 563 704, width=2.4 in]{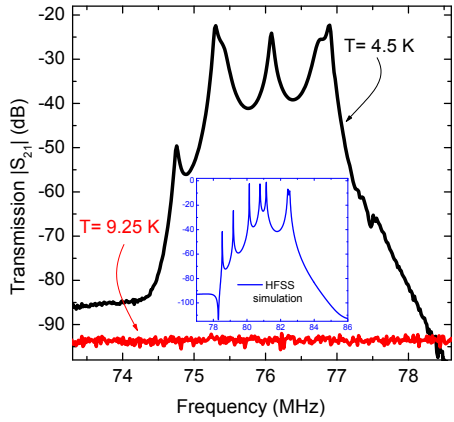}
\caption{(Color online) Main panel: Measured transmission $|S_{21}|$
vs. frequency spectrum on an array of 7 spirals with $D_o$ = 6 mm, N
= 40, $w$ = $s$ = 10 $\upmu$m below and above the $T_c$ of Nb with
an input RF power of $P$ = 0 dBm. Inset: The calculated tranmission
on the same array by HFSS.} \label{Fig6}
\end{figure}

\begin{figure}
\centering
\includegraphics[bb=12 232 593 713, width=2.4 in]{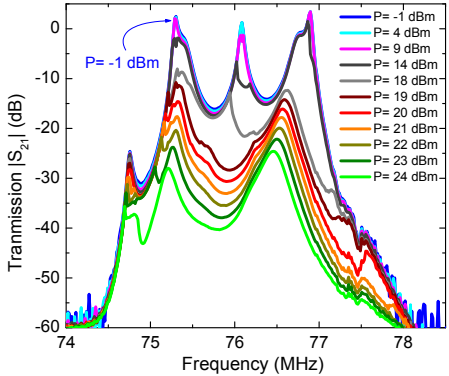}
\caption{(Color online) Measured transmission $|S_{21}|$ vs.
frequency on the same array shown in Fig. 6 at various RF input
powers ranging from $P$ = -1 dBm to $P$ = 24 dBm. Increasing RF
input power leads to non-linear features and unstable jumps in the
transmission. Because an amplifier was used for these measurements,
the background level increased compared to that seen in Fig. 6.}
\label{Fig7}
\end{figure}

Fig. 7 shows the measured transmission data on the same 1-D array in
Fig. 6 for a set of RF input powers, $P$ at a fixed ambient
temperature of $\sim$4.2 K. At $P$ = -1 dBm, the resonance peaks are
sharp and have high $Q_L$ (linear case). With increasing input power
dissipative processes nucleate at discrete sites in the spiral and
increase the resistance $R$ of the circuit, which yields nonlinear
resonant features in the transmission spectrum.

\section{Conclusion}
The linear and non-linear transmission characteristics of compact RF
metamaterials employing Nb thin films are reported. In these RF
metamaterials, the ohmic and dielectric losses are minimized by
carrying out the experiments below the $T_c$ of Nb. The 2-D design
simplifies the fabrication and maintains the uniformity of the
artificial constituent elements. The robust magnetic response of the
spirals can be sensitively tuned by manipulating the superconducting
phase either with temperature or RF magnetic field. The resonant
features in the spectra have an extremely high $Q_L$ (compared to
those of alternative normal metal metamaterials), making them
promising for RF applications.


\end{document}